\documentclass{aa}
\usepackage{graphicx}
\begin{document}
\title{HDO abundance in the envelope of the solar-type
protostar IRAS 16293-2422
\thanks{Based on observations with the IRAM 30m telescope in Spain and with The
James Clerk Maxwell Telescope, operated by The Joint Astronomy Centre on behalf
of the Particle Physics and Astronomy Research Council of the United Kingdom,
the Netherlands Organisation of Scientific Research, and the National Research
Council of Canada.}}

      \subtitle{}
      \titlerunning{HDO abundance in IRAS16293$-$2422}
      \author{B. Parise\inst{1}
             \and E. Caux\inst{1}
             \and A. Castets\inst{2}
             \and C. Ceccarelli\inst{3}
             \and L. Loinard\inst{4}
             \and A.G.G.M. Tielens\inst{5}
             \and A. Bacmann\inst{2}
             \and S. Cazaux\inst{6}
             \and C. Comito\inst{7}
             \and F. Helmich\inst{5,8}
             \and C. Kahane\inst{3}
             \and P. Schilke\inst{7}
             \and E. van Dishoeck\inst{9}
              \and V. Wakelam\inst{2}
             \and A. Walters\inst{1}
             }

      \offprints{B. Parise}

      \institute{Centre d'Etude Spatiale des Rayonnements, BP 4346, F-31028
Toulouse Cedex 04,  \email{parise@cesr.fr}
            \and  Observatoire de Bordeaux, BP 89, F-33270 Floirac
            \and Laboratoire d'Astrophysique de l'Observatoire de
Grenoble, BP 53, F-38041
Grenoble Cedex 9
            \and  Centro de Radioastronom\'{\i}a y Astrof\'{\i}sica, Universidad Nacional Aut\'onoma de
M\'exico, Apartado Postal 72-3 (Xangari), 58089 Morelia, Michoac\'an, Mexico
            \and Kapteyn Astronomical Institute, University of Groningen, P.O. Box 800, 9700 AV Groningen, The Netherlands
            \and INAF, Osservatorio Astrofisico di Arcetri, Largo Enrico
Fermi 5, 50125 Firenze, Italy
            \and Max-Planck-Institut f\"ur Radioastronomie, Auf dem
H\"ugel 69, 53121
Bonn, Germany
            \and SRON National Institute for Space Research, Landleven 12, 9747 AD Groningen, The Netherlands
            \and Leiden Observatory, PO Box 9513, 2300 RA Leiden, The
Netherlands}

      \date{Received June xx, 2004; accepted July yy, 2004}

      \abstract{ We present IRAM 30\,m and JCMT observations of HDO 
lines towards the solar-type protostar IRAS 16293$-$2422. Five HDO transitions
have been detected on-source, and two were unfruitfully searched for towards a 
bright spot of the outflow of IRAS 16293$-$2422. We interpret the data by means
of the Ceccarelli, Hollenbach and Tielens (1996) model, and derive the HDO 
abundance in the warm inner and cold outer parts of the envelope. The emission
is well explained by a jump model, with an inner abundance
x${^{\tiny\rm{HDO}}_{\rm in}}$\,=\,1$\times$10$^{-7}$ and an outer abundance
x${^{\tiny\rm{HDO}}_{\rm out}}\,\leq\,1\times10^{-9}$ (3$\sigma$). This result
is in favor of HDO enhancement due to ice evaporation from the grains in the
inner envelope. The deuteration ratio HDO/H$_{2}$O is found to be
f$_{in}$\,=\,3\% and f$_{out}$ $\le$ 0.2\% (3$\sigma$) in the inner and outer
envelope respectively and therefore, the fractionation also undergoes a jump in
the inner part of the envelope. These results are consistent with the formation
of water in the gas phase during the cold prestellar core phase and storage of
the molecules on the grains, but do not explain why observations of H$_{2}$O 
ices consistently derive a H$_{2}$O ice abundance of several 10$^{-5}$ to 
10$^{-4}$, some two orders of magnitude larger than the gas phase abundance of
water in the hot core around IRAS 16293$-$2422.
       }

      \maketitle
%

\section{Introduction}

The field of molecular deuteration has seen, in recent years, a burst of new
studies, both observational and theoretical, since the discovery of large amounts
of doubly deuterated formaldehyde (about 10\% with respect to the main
isotopomer) in the low mass protostar IRAS16293$-$2422 (hereinafter
IRAS16293, Ceccarelli et al. 1998, 2001). Following this discovery, other
doubly or triply deuterated molecules have been detected having similarly high D/H
enhancements: ammonia (Roueff et al. 2000; Loinard et al. 2001; van der Tak et
al. 2002; Lis et al. 2002), methanol (Parise et al. 2002, 2004) and hydrogen
sulfide (Vastel et al. 2003).

Triggered by these observations, new models were developed to account for
the large observed D/H molecular ratios (Roberts \& Millar 2000a,b; Rodgers \&
Charnley 2003), with partial success. Nonetheless, it was soon understood that
the key to obtain large molecular deuteration is cold and CO depleted gas, as
confirmed by the observations towards a sample of pre-stellar cores (Bacmann et
al. 2003) and predicted by the afore mentioned models. A step forward in the
comprehension of the deuteration process has been the observation of a very
large amount of H$_2$D$^+$ in the pre-stellar core L1544, where very likely 
H$_2$D$^+$/H$_3^+ \sim 1$ (Caselli et al. 2003), after its first detection in 
the low mass protostar NGC1333 IRAS4A (Stark et al. 1999).
This observational study
triggered new models of gas phase chemistry, which take into account all
deuterated isotopomers of H$_3^+$ (Roberts, Herbst \& Millar 2003; Walmsley,
Flower \& Pineau des For\^ets 2004). The comparison between model predictions
and observations is much improved in this last class of models, also supported by
the recent detection of D$_2$H$^+$ (Vastel, Phillips \& Yoshida, 2004).

Molecules like formaldehyde and methanol are almost certainly grain-surface
products, specifically products of successive CO hydrogenation during the cold
dark cloud phase. When a newly formed star heats up its environment, these
species are released into the gas phase because of the ice mantle evaporation
(Charnley, Tielens \& Millar, 1992, Caselli et al. 1993, Charnley et al. 1997, Tielens
\& Rodgers 1997). Therefore, their large deuteration must also occur on the
grain surfaces (e.g. Ceccarelli et al. 2001; Parise et al. 2002, 2004). Note that
fractionation ratios of 0.3, 0.06 and 0.01 have been measured for CH$_2$DOH,
CHD$_2$OH and CD$_3$OH respectively (Parise et al. 2004), so that one would
naively expect similarly large HDO/H$_2$O ratios if water forms on the grains
simultaneously with methanol. However, searches in low-mass sources where large
D$_2$CO/H$_2$CO ratios have been measured have shown no HDO ices at a very low
limit ($\leq 2$\%; Parise et al. 2003). While early analysis of the ISO-SWS 
spectrum of the high-mass protostars W33A and NGC7538 IRS9 led to  
HDO/H$_2$O$_{\rm ice}$ ratios of respectively 8$\times$10$^{-4}$ and 
10$^{-2}$ (Teixeira et al. 1999), reanalysis of this data and supporting ground-based 
data also derived upper limits of 1\% (Dartois et al. 2003).
One possibility is that the process of
water formation on ices is intrinsically unfavorable to water deuteration
because of the involved routes or, alternatively, it is possible that gas phase
and solid phase observations do not probe the same components (see also the
discussion in Parise et al. 2003). Whatever the answer is, it is clear that the
process of molecular deuteration will not be fully mastered until this last
puzzle has a satisfying solution.

The HDO fractionation has already  been measured in a number of high-mass 
hot cores. The HDO/H$_2$O ratio was observed to be 3$-$6 $\times$ 10$^{-4}$ 
in a sample of galactic hot cores (Jacq et al. 1990). Subsequent observations derived similar
fractionation ratios in other high-mass YSO (Gensheimer et al. 1996, Helmich et al. 1996,
Comito et al. 2003).

In order to address the fundamental question of water versus formaldehyde and methanol deuteration, we carried out observations of
five HDO vapor lines towards the low mass protostar IRAS16293, to measure
the HDO/H$_2$O ratio {\it in the gas phase}, and compare it with the observed
fractionations for formaldehyde (Loinard et al. 2000), and methanol (Parise et
al. 2004). Note that IRAS16293 is one of the few sources where the water
abundance profile has been derived, based on ISO-LWS observations (Ceccarelli
et al. 2000a). Several studies have shown that the envelope of IRAS16293
consists of an outer envelope where the molecular abundances are similar to
molecular cloud ones, and an inner envelope where several species have enhanced
abundances because of grain mantle evaporation (Ceccarelli et al. 2000a, 2000b,
2001; Sch\"oier et al. 2002, 2004; Cazaux et al. 2003). It is worth emphasizing
that, in this respect, IRAS16293 is fully representative of solar-mass Class 0
sources (Maret et al. 2004, J{\o}rgensen et al. 2004). Finally, Stark et al. (2004)
recently reported the detection of the HDO ground transition towards
IRAS16293 and derived a HDO abundance of $\sim 10^{-10}$ in the cold region
of the envelope. These authors report only upper limits of higher-lying HDO
transitions, which prevented an accurate estimate of the HDO abundance in the
warm region. We report here the detection of five HDO lines with energies up to 168\,K,
which allows a study of the HDO abundance in the inner envelope.

The article is organized as follows: the observations and results are presented in section 2, 
the modeling and its uncertainties are described in section 3, and the implications of the results 
are discussed in section 4. 

\section{Observations and results}

\subsection{Observations} \label{observations}

IRAS 16293 is known to be comprised of two components, "A" and "B", separated 
from one another by about 5 arcseconds (Wootten 1989, Mundy et al. 1992).
The observations were performed at the JCMT and at the IRAM 30\,m telescopes on
the IRAS16293\,$``$B" source at $\alpha$(2000.0)\,=\,16$^h$ 32$^m$ 22.6$''$,
$\delta$(2000.0)\,=\,$-$24$^{\circ}$ 28$'$ 33$''$. 
The resolution of the observations reported here is never sufficient to resolve
the binary system. The emission of both components is included in the beam used for the 
observations (10$''$ to 33$''$). Some of these data have been obtained from an unbiased
spectral survey of IRAS16293 conducted at IRAM and JCMT by a European Consortium.

\begin{figure}[!htbp]
\includegraphics[width=8.8cm,angle=0]{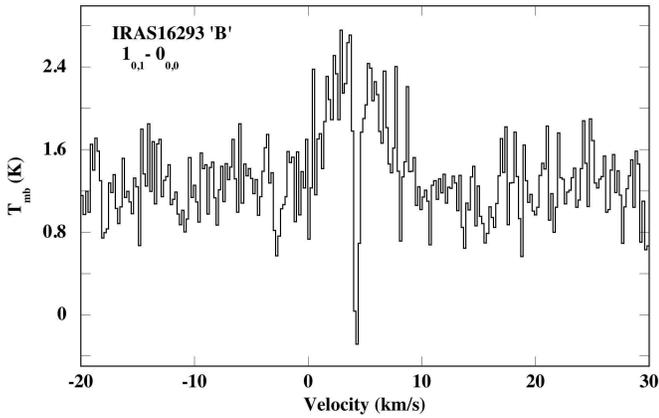}
\caption{HDO 464.9\,GHz line observed on-source (IRAS16293\,``B") at the
JCMT.}
\label{hdo464}
\end{figure}

The ground (1$_{0,1}$--\,0$_{0,0}$) transition of HDO at $\nu$ = 464.9 GHz was
observed on July 26th, 1999 with the JCMT near the summit of Mauna Kea in Hawaii,
USA. The observations were made with the single-sideband dual-polarization W
receiver. Each polarization of the receiver was connected to a unit of an
autocorrelator providing a bandwidth of 250 MHz for a spectral resolution of 156
kHz. At 465 GHz, this yields a velocity resolution of about 0.1 km s$^{-1}$. The
observations were made in position switching mode with the OFF position at
offset $\Delta \alpha =  -180''$, $\Delta \delta = 0''$ from our nominal
position. The spectrum obtained is presented in Fig. \ref{hdo464}. The narrow
self absorption is due to the surrounding cloud (see also Stark et al. 2004).

All other observations were performed with the IRAM 30\,m telescope on Pico
Veleta near Granada, in Southern Spain. To probe where the location of the HDO
emission originates (warm envelope of the source or outflow?), we observed in
addition a position in the flow, at $\Delta\alpha = -39''$,
$\Delta\delta\,=\,0''$ from the on-source nominal position. This position was
chosen, first because it is the location of one of the brightest emissions of the
outflow (CO, Stark et al., 2004), and second to make sure that we do not
intercept emission from the warm envelope of the protostar in the large 33$''$
beam of the 30\,m at 80.6\,GHz.

For on-source observations, we used the beam-switching observing mode, with a
symmetric switch of $240''$ from the nominal center of the source. For the flow
observations, we used the position-switching observing mode, with a switch of
$\Delta\alpha = -3600''$, $\Delta\delta = 0''$ to ensure a reference position
well outside the outflow. Two receivers were always used simultaneously,
connected to a unit of an autocorrelator or filter bank backend.

\begin{figure}[!htbp]
\includegraphics[width=9cm,angle=-90]{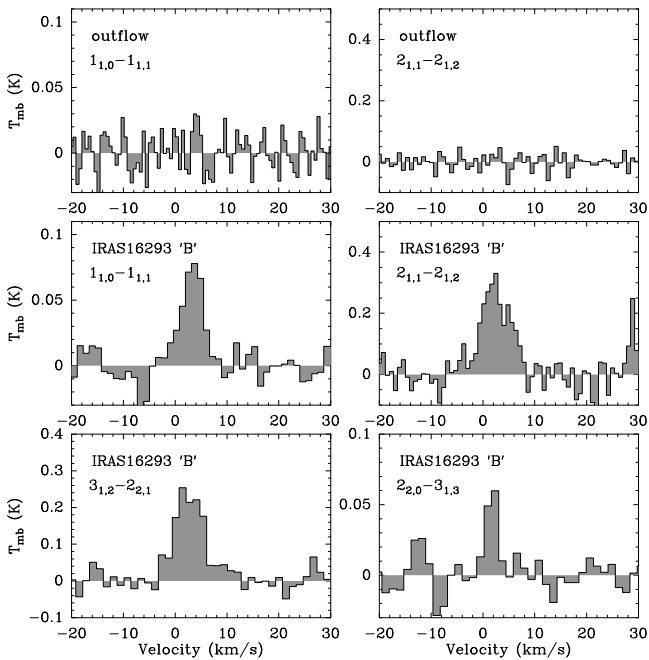}
\caption{HDO lines observed at the 30\,m on the outflow and on-source
(IRAS16293\,``B").}
\label{hdoall}
\end{figure}

The spectral resolutions used, angular resolutions of the telescope, integration
times (ON+OFF) and system temperatures are quoted in Table \ref{table} for both
JCMT and 30\,m observations. Pointing and focus were regularly checked using
planets or strong quasars, providing a pointing accuracy of about $3''$ for both
telescopes.

All intensities reported in this paper are expressed in units of main-beam
brightness temperature, using the efficiencies given on the JCMT and 30\,m web
sites (http://jach.hawaii.edu/JACpublic/JCMT/home.html and 
http://www.iram.fr/IRAMES/index.htm).

\subsection{Results} \label{results}

\begin{figure}[!htbp]
\includegraphics[width=3.2cm,angle=-90]{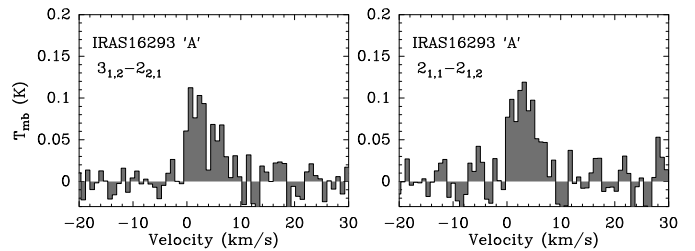}
\caption{Rereduction of the HDO 225.9 and 241.6\,GHz lines observed by Stark et al. in 
2001 at JCMT on IRAS16293\,``A''.}
\label{hdostark}
\end{figure}

The obtained spectra are presented in Figs. \ref{hdo464} and \ref{hdoall} and
show that on the flow position the two searched lines are not detected at all
while all observed lines are detected on-source. The intensity of the
HDO ground transition at 464.9 GHz is very similar to what Stark et al. (2004)
observed at a position centered on IRAS16293\,``A", 5$''$ away from our
IRAS16293\,``B'' position, where they find an integrated flux about
10\% larger than ours. This is not the case for the 225.9 and 241.6 GHz lines,
for  which Stark et al. (2004) reported very low upper limits ($\le$ 120 mK\,km/s
assuming a 6 km/s linewidth). We retrieved from the JCMT database the original
observations performed by Stark on the 225.9 and 241.6 GHz lines and reduced the
data again. The results are shown in Fig.\ref{hdostark}, where the two HDO lines
are clearly seen at the 100\,mK level, which is in good agreement with our
result taking into account the beam dilution in the JCMT telescope. Our results
are also in good agreement with the observation of the 241.6 GHz line reported
by van Dishoeck et al. (1995).

Table \ref{table} summarizes the results of all the observational sets.
Because of the presence of an absorption component, which is obvious for the
ground transition and may be present for other lines, we defined the
integrated intensity for all lines as the sum of all channels in the velocity
range [-5, 10]. The quoted linewidths are those of a Gaussian fit 
to the data. The $\delta \nu$ is the spectral resolution obtained after Hanning
windowing (if any) in Figs. 1 to 3.

\begin{table*}[!htbp]
\begin{center}
\vspace{0.5cm}
\caption{HDO lines parameters. The three sections are for the three aimed
positions. In italics, we quoted the results published by Stark et al. (2004).}
\label{table}
\begin{tabular}{ccccccccccccccc}
\hline \hline
\noalign{\smallskip}
Observing & Telescope &Transition  & Frequency &  Eup & Beam & T$_{int} $ &
T$_{sys}$  & $\delta$v & RMS & T$_{peak}$ & $\Delta$v &
$\int$T$_{mb}$dv  \\
Date &  & & GHz & K & $''$ & min  & K  & km/s & mK & mK &
km/s & K.km/s\\
\noalign{\smallskip}
\hline
\noalign{\medskip}
\multicolumn{13}{c}{IRAS16293\,``B"
($\alpha$(2000.0)\,=\,16$^h$ 32$^m$ 22.6$''$,
$\delta$(2000.0)\,=\,$-$24$^{\circ}$ 28$'$ 33$''$)} \\
\noalign{\smallskip}
\hline
\noalign{\smallskip}
01/20/04 & IRAM &1$_{1,0}$--1$_{1,1}$ & 80.578 & 46.8 & 33 & 42 &190  & 1.2
& 10 & 81 & 4.9 & 0.40 \\
01/25/04 & IRAM & 3$_{1,2}$--2$_{2,1}$ & 225.897 & 167.7 &12 & 60 &650  & 1.3
& 12 & 245 & 6.2 & 1.7  \\
03/29/00 & IRAM & 2$_{1,1}$--2$_{1,2}$ & 241.561 & 95.3 &11 & 35 &880  & 0.4
& 62 & 400 & 6.6 & 2.0  \\
02/03/04 & IRAM & 2$_{2,0}$--3$_{1,3}$ & 266.161 & 157.3 &10 & 68 &820  & 1.4
& 14 & 75 & 3.0 & 0.21  \\
07/26/99 & JCMT & 1$_{0,1}$--0$_{0,0}$  & 464.924 & 22.3 & 11 & 140 & 3000 & 0.2 
& 290 & 1200 & 6.0 & 5.5  \\
\noalign{\smallskip}
\hline
\noalign{\medskip}
\multicolumn{13}{c}{Flow ($\alpha$(2000.0)\,=\,16$^h$ 32$^m$ 20$''$,
$\delta$(2000.0)\,=\,$-$24$^{\circ}$ 28$'$ 33$''$)} \\
\noalign{\smallskip}
\hline
\noalign{\smallskip}
03/09/04 & IRAM & 1$_{1,0}$--1$_{1,1}$ & 80.578 & 46.8 &33 & 40 & 230 & 0.58 &
14 & - & 6.0$^{*}$  &  $<$ 0.05$^{*}$  \\
03/09/04 & IRAM & 2$_{1,1}$--2$_{1,2}$ & 241.561 & 95.3 &11 & 40 & 950 & 0.4 &
38 & - & 6.0$^{*}$  & $<$ 0.01$^{*}$  \\
\noalign{\smallskip}
\hline
\noalign{\medskip}
\multicolumn{13}{c}{IRAS16293\,``A" ($\alpha$(2000.0)\,=\,16$^h$
32$^m$ 22.85$''$,
$\delta$(2000.0)\,=\,$-$24$^{\circ}$ 28$'$ 35.5$''$)} \\
\noalign{\smallskip}
\hline
\noalign{\smallskip}
\it{1998} & \it{JCMT} & \it{1$_{0,1}$--0$_{0,0}$}  & \it{464.924} & \it{22.3} &
\it{11} & - & - & - & - & \it{1000} & \it{5.9} & \it{6.0}  \\
2001 & JCMT & 3$_{1,2}$--2$_{2,1}$ & 225.897 & 167.7 & 21 & 120 & 715  & 0.83
& 18 & 87 & 6.5 & 0.61  \\
2001 & JCMT & 2$_{1,1}$--2$_{1,2}$ & 241.561 & 95.3 & 20 & 96 & 690 & 0.78
& 23 & 112 & 5.4 & 0.62  \\
\noalign{\smallskip}
\hline
\end{tabular}
\end{center}
$^{*}$: For the outflow observations, since no linewidth can be determined, a
mean width of 6 km/s was assumed. The data have been smoothed to the velocity
resolution $\delta$v given in column 8, used to determine the RMS noise. ``-''
means the information is not available or cannot be derived. The last 3 observations are 
from Stark et al. (2004).
\end{table*}

Except for the 266.2 GHz line, which is the noisiest one, the observed linewidths
are broad -- $\sim$ 6 km/s -- and therefore should come mainly from either the
infalling inner warm envelope or from the outflow, rather than from the
cold envelope. Furthermore, the observed intensity for both 225.9 and
241.6\,GHz HDO lines is very different at JCMT and at the 30\,m. This can be
explained if the emission of these lines comes from a very small region, more
diluted in the JCMT beam than it is in the 30\,m beam. If we assume the size of
the emitting region to be small with respect to the 30\,m beam (Ceccarelli
et al. 2000a modelled  2$''$), we expect a flux about 4 times larger in the
30\,m beam, very similar to what we observe (2.8 at 225.9\,GHz and 3.2 at
241.6\,GHz). We attribute the residual disagreement to slightly different
positions between the Stark (IRAS16293\,``A'') and our (IRAS16293\,``B'')
observations.

If the HDO emission arises from a very small region, this argues in favour of
the warm envelope for the origin of the emission, rather than from the outflow.
This is also strongly suggested by the non-detection of both 80.6 and 
241.6\,GHz lines towards the outflow at the 30\,m. We will therefore model the 
observed HDO line emission assuming the lines originate in the envelope of the
protostar.

\section{Modeling and discussion}

\subsection{Modeling}

The structure of the envelope of IRAS16293 was derived by Ceccarelli et al.
(2000a) using H$_2$O lines observed with ISO-SWS and ISO-LWS, and
substantially confirmed by the subsequent analysis of Sch\"oier et al. (2002). The
water emission was modeled in terms of a jump model (Ceccarelli Hollenbach and
Tielens 1996, hereinafter CHT96), where the abundances of water in the
inner part of the envelope (T$\,\ge\,$100\,K, evaporation temperature of the icy
grain mantles) and in the outer part (T$\,\le\,$100\,K) are two free parameters.
The derived inner abundance was x${^{\tiny\rm{H_{2}O}}_{\rm in}}$\,=\,3$\times
10^{-6}$ (with respect to H$_2$) and the outer abundance  x${^{\tiny\rm{H_{2}O}}_{\rm
out}}$\,=\,5$\times10^{-7}$ (Ceccarelli et al. 2000a).

Studies of the spatial distribution of formaldehyde in IRAS16293 
have shown that the structure may be more complex than a single step function, 
as a further jump may be present at around 50 K, due to evaporation of CO-rich ices
(Ceccarelli et al. 2001, Sch\"oier et al. 2004). Given the low number of observed 
transitions, we will consider here the simple case of a single jump. The abundance 
derived in the outer region will therefore likely be an average over the regions where
CO is depleted and starts to evaporate. 

\begin{figure}
\includegraphics[width=9.4cm]{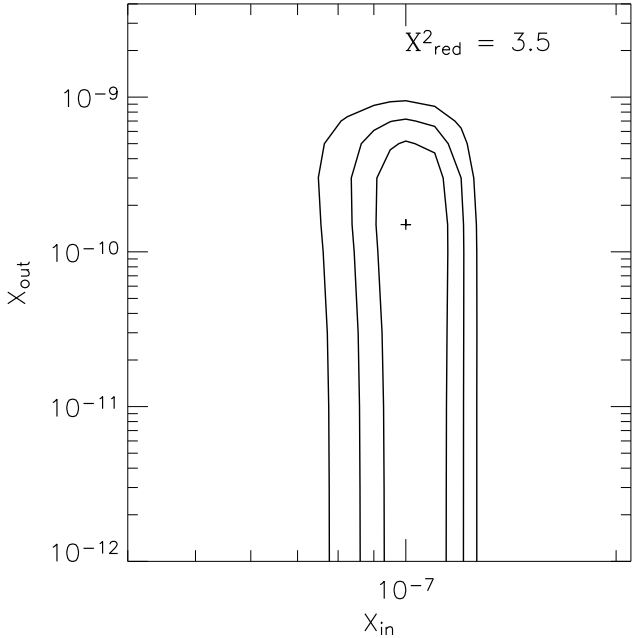}
\caption{x${^{\tiny\rm{HDO}}_{\rm in}}$ and x${^{\tiny\rm{HDO}}_{\rm out}}$
contours (1, 2 and 3$\sigma$) for the reduced $\chi^2$. The ``+'' corresponds to
the best fit model. x${^{\tiny\rm{HDO}}_{\rm in}}$ is very well constrained, 
($1\pm 0.3\times10^{-7}$), as well as the upper limit of 
x${^{\tiny\rm{HDO}}_{\rm out}}$ ($\leq 1\times10^{-9}$, 3$\sigma$).}
\label{ki2}
\end{figure}

\begin{figure}[!htbp]
\includegraphics[width=9cm,angle=0]{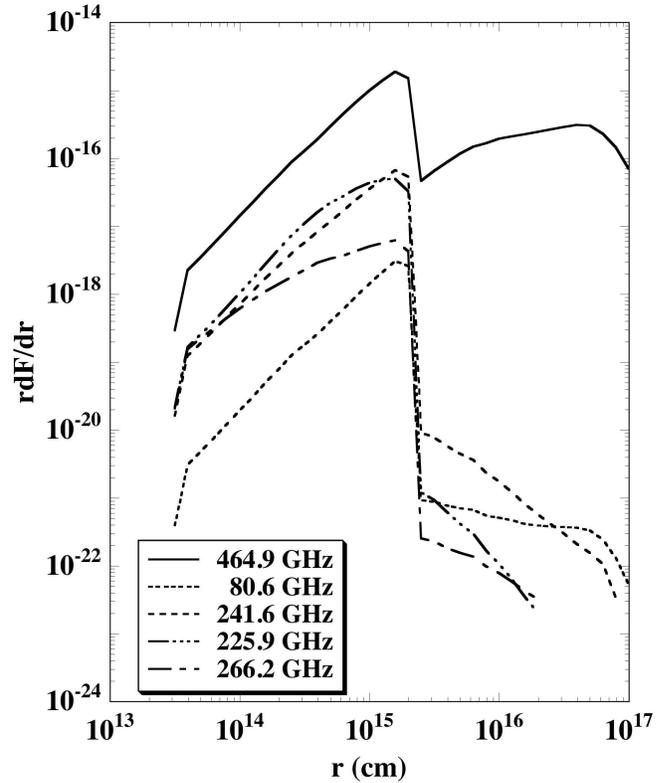}
\caption{Radial emission profiles of the five HDO lines, 
using x${^{\tiny\rm{HDO}}_{\rm in}}$\,=\,1$\times$10$^{-7}$ and
x${^{\tiny\rm{HDO}}_{\rm out}}\,=\,1.5\times$10$^{-10}$.}
\label{radial-profile}
\end{figure}

For the analysis of the present HDO data, we adapted the time-dependent 
CHT96 model to
compute the HDO line emission at a given time. The collisional coefficients were taken
from Green et al. (1989), and the details of the model are reported in Parise,
Ceccarelli \& Maret (2004). We adopted the temperature and density structure
derived by Ceccarelli et al. (2000a) for the envelope and left the inner and
outer HDO abundances as free parameters. We then performed a $\chi^2$ analysis
for x${^{\rm \tiny{HDO}}_{\rm in}}$ ranging from 1$\times$10$^{-9}$ to
1$\times$10$^{-6}$ and for x${^{\tiny\rm{HDO}}_{\rm out}}$ ranging from
1$\times$10$^{-12}$ to 1$\times$10$^{-8}$. The best model fitting the
5 observed lines on-source corresponds to x${^{\tiny\rm{HDO}}_{\rm
in}}$\,=\,1$\times$10$^{-7}$ and x${^{\tiny\rm{HDO}}_{\rm
out}}\,=\,1.5\times$10$^{-10}$, and gives a reduced $\chi^2$ of 3.5. Figure
\ref{ki2} presents the contours delimitating the 1$\sigma$, 2$\sigma$ and
3$\sigma$ confidence intervals (corresponding respectively to ${\rm
\chi^2_{red}}$\,=\,${\rm \chi^2_{min}}$+1.18,
${\rm \chi^2_{min}}$+2.70 and ${\rm \chi^2_{min}}$+5.06 as relevant
for 3 degrees of freedom). The inner abundance is very well constrained, while
the data only provide an upper limit on the outer abundance. The lower limit on
the outer abundance is poorly constrained, because the only transition
constraining it is the ground transition at 464.9 GHz. Fig \ref{radial-profile}
shows the radial profile of the emission of the five HDO lines computed with 
x${^{\tiny\rm{HDO}}_{\rm in}}$\,=\,1$\times$10$^{-7}$ and 
x${^{\tiny\rm{HDO}}_{\rm out}}\,=\,1.5\times$10$^{-10}$. It is clear on 
this figure that only the ground transition has a contribution from the outer
envelope, and even more that the bulk of the emission originates in the inner
part of the envelope.

We also performed the same analysis with only the 3 lines observed on
IRAS16293\,``A" at JCMT (225.9, 241.6 and 464.9 GHz). The resulting abundances
are x${^{\tiny\rm{HDO}}_{\rm in}}$\,=\,1.1$\times$10$^{-7}$ and
x${^{\tiny\rm{HDO}}_{\rm out}}$\,$\leq$\,1$\times$10$^{-9}$ (3$\sigma$), compatible
with the results found on IRAS16293\,``B". Note that with their analysis, Stark
et al. (2004) estimate a constant HDO abundance of 3$\times$10$^{-10}$
throughout the envelope, compatible with the abundance we derive in the outer
envelope. On the contrary, they do not find an abundance jump in the warm inner
envelope, presumably because they only used the ground transition to
constrain it.

\subsection{Uncertainties of the model}

Following the discussion in Maret et al. (2004), the values of the inner and
outer abundances derived by our model can be uncertain for several reasons
that we review below:

\smallskip
- To test the influence of the outer abundance in the derivation
of the inner abundance, we arbitrarily imposed an outer abundance one order of
magnitude greater than the derived abundance (simulating e.g. an extreme
absorption of the ground transition by foreground clouds). We then constrained
the inner abundance without using the ground transition. The best fit is
still obtained for the same value of x${^{\tiny\rm{HDO}}_{\rm
in}}$ and we can thus conclude that this result is robust regardless of any
foreground absorption of the ground transition.

\smallskip
- To check the validity of the jump model, we ran a model with a constant
HDO abundance throughout the envelope. The best fit is obtained in this case
for an abundance of 1.2$\times$10$^{-9}$, but the fit is very poor, yielding
a reduced $\chi^2$ of 40. We conclude from this analysis that a jump model
is required to account for the observed HDO emission.

\smallskip
- In order to test the influence of the evaporation temperature (assumed to be
100\,K in the present study), we also ran the model for an evaporation
temperature of 50\,K. The model using this new input parameter poorly fits our
observations. Indeed, the best fit is obtained with a reduced $\chi^2$
of 42 to be compared to the value of 3.5 when the evaporation temperature
is 100\,K. This analysis is in good agreement with the measured evaporation
temperature of water-rich ices (Sandford \& Allamandola, 1990).

\smallskip
- As noted previously, the 464.9 GHz linewidth is about 6 km/s, which would
suggest that it originates from the inner warm region. To check if this is true, 
we ran the model with a very low value of the outer HDO abundance, 
x${^{\tiny\rm{HDO}}_{\rm out}}$\,= \,7.5$\times$10$^{-12}$, i.e. with no
enhancement with respect to the cosmic abundance (D/H$_{{\rm ISM}}$ = $1.5\times
10 ^{-5}$, Linsky et al. 1998) and using x${^{\tiny\rm{H_{2}O}}_{\rm
out}}$\,=\,5$\times$\,10$^{-7}$, (Ceccarelli et al. 2000a). In this last case,
the best fit corresponds to x${^{\tiny\rm{HDO}}_{\rm in}}$\,=
\,1.05$\times$10$^{-7}$, the bulk of the ground HDO transition originates in
the inner region, and the model underestimates the observed flux by only
15$\%$. Therefore, the 6 km/s linewidth of the ground HDO transition is
consistent with our model as most of it originates in the inner warm region.
Of course, the presence of the narrow self-absorption feature suggests that 
while most of the 464.9\,GHz emission originates from the warm inner envelope, some 
HDO has to be present in the outer cold, absorbing envelope.

\begin{figure}[!htbp]
\includegraphics[width=8.8cm,angle=0]{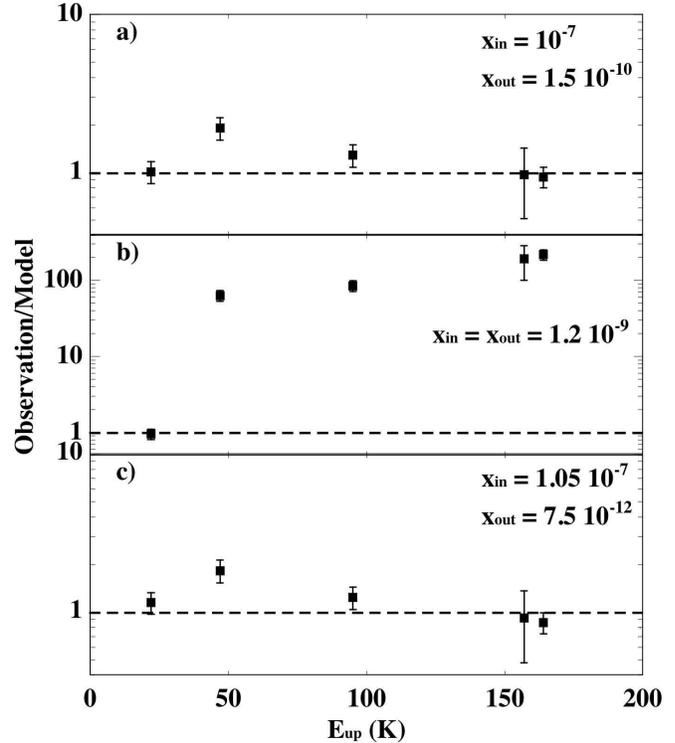}
\caption{Ratios between the observations on IRAS16293\,``B'' and the model
predictions for three cases : a) the jump model with 
x${^{\tiny\rm{HDO}}_{\rm in}}$\,=\,1$\times$10$^{-7}$
and x${^{\tiny\rm{HDO}}_{\rm out}}\,=\,1.5\times$10$^{-10}$, b) a model with a
constant abundance throughout the envelope, and c) a case of no enhancement at
all of the emission associated with the outer envelope (see text).}
\label{obs-models}
\end{figure}

\smallskip
Thus, Fig. \ref{obs-models} shows the ratios between the observations on
IRAS16293\,``B'' and the model predictions for three cases : a) the jump
model with x${^{\tiny\rm{HDO}}_{\rm in}}$\,=\,1$\times$10$^{-7}$
and x${^{\tiny\rm{HDO}}_{\rm out}}\,=\,1.5\times$10$^{-10}$,
b) the case with a constant abundance throughout the envelope
(x${^{\tiny\rm{HDO}}_{\rm in}}$\,=\,x${^{\tiny\rm{HDO}}_{\rm
out}}$\,=\,1.2$\times$10$^{-9}$), and c) the case where the HDO abundance in the
outer envelope is x${^{\tiny\rm{HDO}}_{\rm out}}$\,=
\,7.5$\times$10$^{-12}$.

All the checks done strengthen the fact that the observations are consistent with
the previously derived HDO inner and outer abundances. They are summarized in Table 
\ref{summary}. These values lead, when compared to
the H$_2$O abundances determined by Ceccarelli et al. (2000a),  
to the fractionation ratios indicated in Table \ref{summary}. Note that these
H$_2$O abundances are also relatively uncertain. In particular, the inner abundance could be 
underestimated as it is derived from optically thick lines. Although Ceccarelli et al. (2000a) 
provide an upper limit on x${^{\tiny\rm{H_2O}}_{\rm in}}$ of 3.5$\times$10$^{-6}$, 
future observations of water lines with the Herschel-HIFI spectrometer are needed to reduce 
the uncertainties on the water distribution. 

\begin{table}[!htbp]
\begin{center}
\caption{Summary of the results of the modelling.}
\begin{tabular}{ccc}
\hline
\hline
\noalign{\smallskip}
& inner envelope & outer envelope \\
\noalign{\smallskip}
\hline
\noalign{\smallskip}
x${^{\tiny\rm{HDO}}}$ & 1$\times$10$^{-7}$  &  $\leq\,1\times$10$^{-9}$ \\
x${^{\tiny\rm{H_2O}}}^*$ & 3$\times 10^{-6}$ & 5$\times10^{-7}$ \\
HDO/H$_2$O & 3\% & $\le$ 0.2\% (3$\sigma$) \\
\noalign{\smallskip}
\hline
\noalign{\smallskip}
\end{tabular}
\label{summary}
\end{center}
$^*$: Ceccarelli et al. 2000a. 
\end{table}

\subsection{Discussion}

These results clearly show that the abundance of HDO undergoes a jump in
the inner part of the envelope, where the ices evaporate from the grains,
and that, even more strikingly, the fractionation also undergoes such
a jump. This is not in agreement with the results of Stark et al. (2004), who
found an equal HDO abundance in the inner and outer envelope of the source and 
a HDO/H$_{2}$O ratio of 0.15\% in the inner warm envelope and 2 to 20\% in the
outflow. Regarding the abundance in the inner and in the outer envelope, our
analysis of several lines demonstrates that indeed there is a region where the
HDO abundance exhibits a jump. On the contrary, we do not have any observational
evidence that HDO is associated with the outflow as we do not detect any emission
in the position of the outflow (see Fig. \ref{hdoall} and Table
\ref{table}). However, we cannot totally rule out that at least part of the HDO 
emission comes from an interaction of the envelope with the outflow, as suggested by 
Stark et al. (2004).

The deuteration fractionation of water derived in the inner part of the
envelope is lower by one order of magnitude than the fractionation of methanol
(30\% for CH$_2$DOH, Parise et al. 2002, 2004) and formaldehyde 
(15\%, Loinard et al. 2000). This result is consistent with the non detection of
solid HDO towards low-mass protostars which exhibit a high deuteration of
formaldehyde in the gas phase (Parise et al. 2003). The present analysis confirms
that water is indeed less deuterated than formaldehyde and methanol in the hot
core of low-mass protostars.

Comito et al. (2003) derived a fractionation of  6.4\,$\times$\,10$^{-4}$ in
the hot core region of the SgrB2 complex, similar to the water fractionation
HDO/H$_{2}$O = (2-6)\,$\times$\,10$^{-4}$ found in the high-mass protostar W3 by
Helmich, van Dishoeck \& Jansen (1996). Such low values of the HDO fractionation
(a few 10$^{-4}$) have also been derived in some high-mass star forming regions
by the pioneering work of Jacq et al. (1990). Our results show that the water
fractionation in the solar-type protostar IRAS16293 is much higher than what is
observed in high-mass protostars, as already pointed out for the formaldehyde 
(Loinard et al. 2002) and methanol fractionation (Jacq et al. 1993; Parise et al.
2002, 2004).

\smallskip
The jump by more than a factor of 10 in the fractionation of water in the region
where mantles evaporate suggests that the fractionation processes are
substantially different in the two regions: 
\smallskip\\
$\small{\bullet}$ In the outer envelope, where the dust temperature is not high enough to
efficiently evaporate the molecules stored in grain mantles, the fractionation
might reflect current gas-phase deuteration processes. In the gas-phase scheme,
the deuteration is driven by reactions with H$_2$D$^+$. This can lead to a water
fractionation enhancement of up to several percent when the temperature is very low
 (T $\sim$ 10\,K, Roberts et al. 2000b; Roberts et al. 2004), because of the
endothermicity of the reaction ${\rm H_2D^+ + H_2 \rightarrow H_3^+ + HD}$,
enhancing the H$_2$D$^+$/H$_3^+$ ratio relatively to the HD/H$_2$ ratio.
In the outer envelope, temperatures span from $\sim$ 10\,K to 100\,K, and the
measured fractionation is thus characteristic of a medium warmer than 10\,K, for which 
fractionation drops very quickly with respect to 10\,K (cf. Fig 2b of Roberts et
al. 2000b). The fractionation value ($\le$ 0.2\%) that we derive is thus in
agreement with this gas phase scheme.
\smallskip\\
$\small{\bullet}$ On the contrary, in the inner envelope, the fractionation may probe the
deuteration of the molecules formed during an earlier cold phase when CO
depletion was extreme, as observed presently in some prestellar cores (Caselli et
al. 1999; Bacmann et al. 2002, 2003; Crapsi et al. 2004). These molecules are
stored in the grain mantles that evaporate once the protostar heats its
surroundings.

\smallskip
In the inner envelope, the difference between the fractionation of water on the
one hand and formaldehyde and methanol on the other hand is, as discussed by
Parise et al. (2003), a strong constraint to chemical models. Because of the low 
efficiency of its production in the gas phase and in view of its high abundance
in icy mantles, methanol is believed to be formed on the grains by active grain
surface chemistry, and successive hydrogenations of CO (Tielens 1983; Charnley,
Tielens \& Millar, 1992; Charnley, Tielens \& Rodgers 1997). If water is also
produced by active grain chemistry, the lower fractionation of water compared
with methanol suggests that either there is a selective incorporation of deuterium in
the methanol route (successive hydrogenations/deuterations of CO, resulting in
the production of formaldehyde and methanol) rather than in the water route, or
water is not formed simultaneously with methanol. 

Such segregation of ices is indicated by solid CO observations towards a sample
of low-mass protostars showing evidence that 60\,$\%$ to 90\,$\%$ of solid CO is
in the form of pure CO-ice (Tielens et al. 1991, Boogert et al. 2002, Pontopiddan
et al. 2003). Likewise, observations of solid CO$_{2}$ also provide evidence for
separate ice components along the same line of sight, although, in this case,
this is generally attributed to the segregation of mixed H$_{2}$O\,/\,CH$_{3}$OH\,/\,CO$_{2}$
ices upon warm up by a newly formed star (Ehrenfreund et al. 1998; 1999;
Gerakines, Moore \& Hudson 2000; Boogert et al. 2000).

Perhaps the water ice observation refers to a global property of molecular clouds while the methanol-rich ices are more localized to regions of star formation. Indeed, studies of the ice abundance suggest that H$_2$O-ice appear wherever A$_{\rm V}$ $>$ 3 magnitudes (Whittet et al. 1988, Chiar et al. 1995), while methanol ice is rarely seen in dark clouds (Chiar et al. 1996).

\smallskip
One of the possibilities discussed by Parise et al. (2003) can be ruled out by
these new observations. Indeed, the possibility that H$_2$O is condensed out
on the grains after a shock during the cloud phase (as suggested by Bergin,
Neufeld \& Melnick, 1999) can be rejected in the case of IRAS16293 as the
deuteration in such a scheme would be lower than a few 10$^{-3}$, i.e. at least
10 times smaller than the fractionation we derive in the inner warm envelope.

\smallskip
Another possibility is that water is produced in the gas phase at low temperature
during the prestellar core phase before it is stored in the grain mantles. The
gas phase model predictions of Roberts et al. (2000b) seem to be in agreement
with this scheme. Indeed, the water fractionation is expected to reach a few
percent in a gas at 10\,K and density n$\,=\,5\times 10^{4}$\,cm$^{-3}$, even
without considering CO depletion (see Fig. 3 of Roberts et al. 2000b). The
water abundance is predicted to be nearly $10^{-6}$ in this case, i.e. only a
factor of 3 below the abundance x${^{\tiny\rm{H_{2}O}}_{\rm in}}$ derived by
Ceccarelli et al. (2000a). Both H$_{2}$O and present HDO observations in the
warm inner envelope may thus be consistent with the formation of water in
the gas phase, the dust playing only a passive role in maintaining the
fractionation at its cold value during storage of the molecules.

While such a model would be consistent with our gas phase observations of
H$_{2}$O and HDO (eg., absolute abundance as well as fractionation behavior),
observations of ices consistently derive a H$_{2}$O ice abundance of 10$^{-4}$
in high-mass protostars (Whittet et al., 1988; Smith, Sellgren \& Tokunaga,
1989; Gibb et al. 2004), and 5$\times$10$^{-5}$ in low-mass protostars (Boogert
et al. 2004), at least one order of magnitude larger than the gas phase abundance
of H$_{2}$O in the hot core around IRAS16293. Such high abundances of H$_{2}$O
ice are generally thought to reflect active grain surface chemistry, eg.
hydrogenation of atomic oxygen on grain surface (Tielens \& Hagen 1982; Jones, Duley \&
Williams 1990). This discrepancy between the hot core H$_{2}$O
abundance in IRAS16293 and the general H$_{2}$O ice abundance may merely reflect
a unique situation for this source but that solution is not very satisfactory. In
particular, IRAS16293 is often considered to be the template solar-type class 0
protostar and, indeed, it shares many properties of class 0 sources (e.g. Ceccarelli
et al. 2000b, Maret et al. 2004). In a way, all models $-$ including the grain
surface chemistry origin of H$_{2}$O $-$ have to face this same problem
of the difference in the hot core and solid state H$_{2}$O abundance. If the gas
phase composition of hot cores really reflects the evaporation of ices, the
H$_{2}$O abundance would be expected to be much higher.
The much lower gaseous H$_2$O abundance in the hot core -- as compared to the H$_2$O-ice  abundance towards protostars -- was already noted by Ceccarelli et al. (2000a). They attributed this discrepancy to a breakdown of spherical symmetry when the size approaches the core-rotation radius ($\sim$\,30\,AU) and the presence of a disk. In this disk, much of the water may be frozen out. At the same time, the disk is also not accounted for in the studies of the total gas column density.
Likely, the HDO/H$_2$O ratio in the inner part is less sensitive to these uncertainties. The HIFI heterodyne instrument on Herschel will provide further insight into these issues.

\section{Conclusion}

Five HDO lines have been detected towards the solar-type protostar
IRAS16293``B'' using the IRAM 30\,m and JCMT telescopes. Two lines (80.6
and 241.6 GHz) were unfruitfully searched for at the 30\,m towards a 
bright spot of the outflow of IRAS16293.

We modeled the emission on-source with the CTH96 jump model, and 
derived the HDO abundance in the inner and outer parts of the envelope to be 
x${^{\tiny\rm{HDO}}_{\rm  in}}$\,=\,1$\times$10$^{-7}$ and x${^{\tiny\rm{HDO}}_{\rm
out}}\,\leq\,1\times10^{-9}$, in agreement with HDO enhancement
due to the ices' evaporation from the grains in the inner envelope.

The water fractionation also undergoes a jump as we obtained f$_{in}$\,=\,3\,\%
and f$_{out}$ $\le$ 0.2\,\% in the inner and outer envelope, respectively. These
results are consistent with the formation of water in the gas phase during the
cold prestellar core phase and storage of the molecules on the grains. They do
not explain why H$_{2}$O observations of ices consistently derive a H$_{2}$O ice
abundance of several 10$^{-5}$ to 10$^{-4}$, some two orders of magnitude larger
than the gas phase abundance of water in the hot core around IRAS16293.

\begin{acknowledgements}

We would like to thank the JCMT and IRAM 
30\,m teams for their hospitality, support and help in the conduction of the
observations. We thank Pierre Valiron for very fruitful discussions that
improved the content of this paper. We thank the referee, Paola Caselli,
for very interesting comments that contributed to improving the paper.
\end{acknowledgements}

\end{document}